# 23.8 GHz Acoustic Filter in Periodically Poled Piezoelectric Film Lithium Niobate with 1.52 dB IL and 19.4% FBW

Sinwoo Cho, Omar Barrera, Jack Kramer, Vakhtang Chulukhadze, Tzu-Hsuan Hsu,
Joshua Campbell, Ian Anderson, *Graduate Student Member, IEEE,* and Ruochen Lu., *Member, IEEE*

*Abstract*—This paper reports the first piezoelectric acoustic filter in periodically poled piezoelectric film (P3F) lithium niobate (LiNbO$_3$) at 23.8 GHz with low insertion loss (IL) of 1.52 dB and 3-dB fractional bandwidth (FBW) of 19.4%. The filter features a compact footprint of 0.64 mm$^2$. The third-order ladder filter is implemented with electrically coupled resonators in 150 nm bi-layer P3F 128° rotated Y-cut LiNbO$_3$ thin film, operating in second-order symmetric (S2) Lamb mode. The record-breaking performance is enabled by the P3F LiNbO$_3$ platform, where piezoelectric thin films of alternating orientations are transferred subsequently, facilitating efficient higher-order Lamb mode operation with simultaneously high quality factor (*Q*) and coupling coefficient (*k$^2$*) at millimeter-wave (mmWave). Also, the multi-layer P3F stack promises smaller footprints and better nonlinearity than single-layer counterparts, thanks to the higher capacitance density and lower thermal resistance. Upon further development, the reported P3F LiNbO$_3$ platform is promising for compact filters at mmWave.

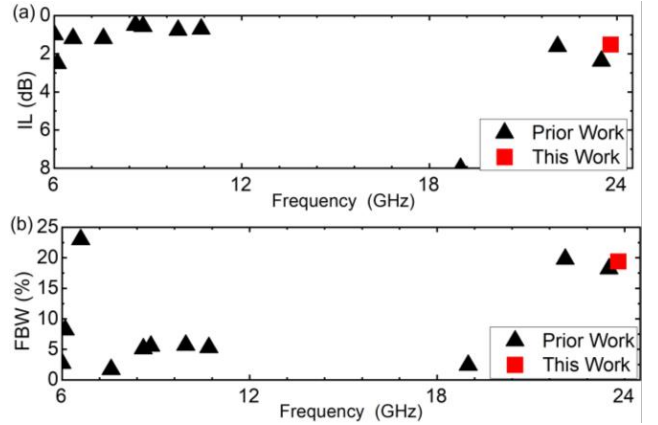

Fig. 1 Survey of (a) IL and (b) FBW in acoustic filters above 6 GHz.

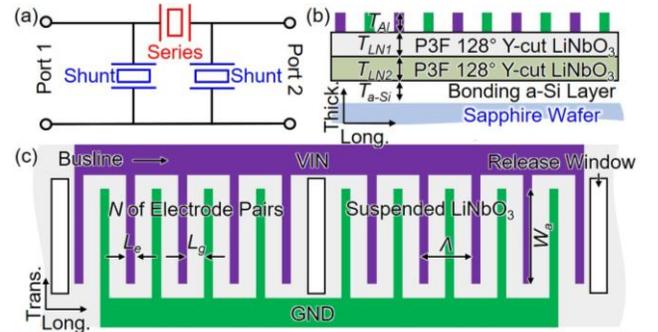

Fig. 2 (a) Filter circuit, and resonator (b) side and (c) top schematics.

## I. INTRODUCTION

The constant pursuit of faster data rates is pushing mobile networks into millimeter-wave (mmWave) frequency bands for wider bandwidth in 5G and beyond wireless systems [1]–[3]. One key research area is compact frequency (RF) front-end components at mmWave [3]. Piezoelectric acoustic devices at RF, where electromagnetic (EM) signals are converted into and processed as mechanical vibrations [4], are good candidates, thanks to the five-order-of-magnitude shorter wavelengths of acoustic waves than EM ones [5]–[7], enabling resonators [8]–[11] and waveguiding structures [12]–[15] with smaller footprints. Because of such advantage, sub-6 GHz front-end filtering has dominantly adopted acoustics, e.g., surface acoustic wave (SAW) devices in lithium niobate (LiNbO$_3$), lithium tantalate (LiTaO$_3$), and thin-film bulk acoustic resonators (FBARs) in aluminum nitride or scandium aluminum nitride (AlN/ScAlN) [16]–[23]. It would be ideal to further leverage acoustics for mmWave RF front ends.

However, it is unclear whether it is feasible to frequency scale application-worthy acoustic filters into mmWave. In conventional platforms, the insertion loss (IL) [Fig. 1(a)] and fractional bandwidth (FBW) [Fig. 1(b)] dramatically degrade beyond 6 GHz [21], [22], [24]–[32]. Scaling conventional acoustic devices has been limited by fabrication challenges from either very thin films or very small lateral feature sizes [33], [34]. Thin piezoelectric films tend to suffer from worse crystalline quality, while narrow electrodes introduce too much resistive loss. New platforms are required.

Recently, first-order antisymmetric mode (A1) resonators in sub-100 nm LiNbO$_3$ on sapphire substrate with intermediate amorphous silicon (a-Si) are demonstrated as low-loss and wideband mmWave platforms [35], [36]. Innovative design and advanced thin-film transfer collectively enable filters at 20 GHz in the state-of-the-art (SoA) (Fig. 1). Despite the great performance, the thin-film filters inherit a few issues. First, the

Manuscript received X February 2024; revised XX February 2024; accepted XX February 2024. This work was supported by DARPA COmpact Front-end Filters at the ElEment-level (COFFEE). S. Cho, O. Barrera, J. Kramer, V. Chulukhadze, J. Campbell, I. Anderson, and R. Lu are with The University of Texas at Austin, Austin, TX, USA (email: sinwoocho@utexas.edu). T.-H. Hsu, is with National Tsing Hua University, Hsinchu, Taiwan; he is currently a visiting student at The University of Texas at Austin, TX, USA. This article was presented at the IEEE MTT-S International Microwave Symposium (IMS 2024), Washington, DC, USA, June 16–21, 2024.

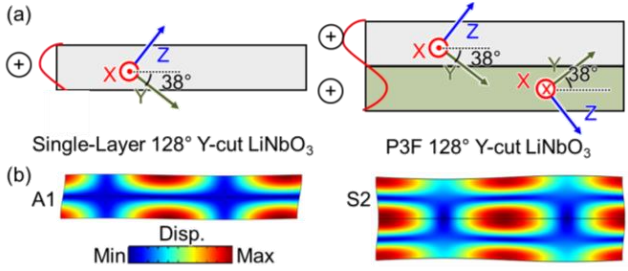

Fig. 3 (a) Side-view of single-layer LiNbO3 and P3F LiNbO3. (b) Displacement mode shape of A1 and S2.

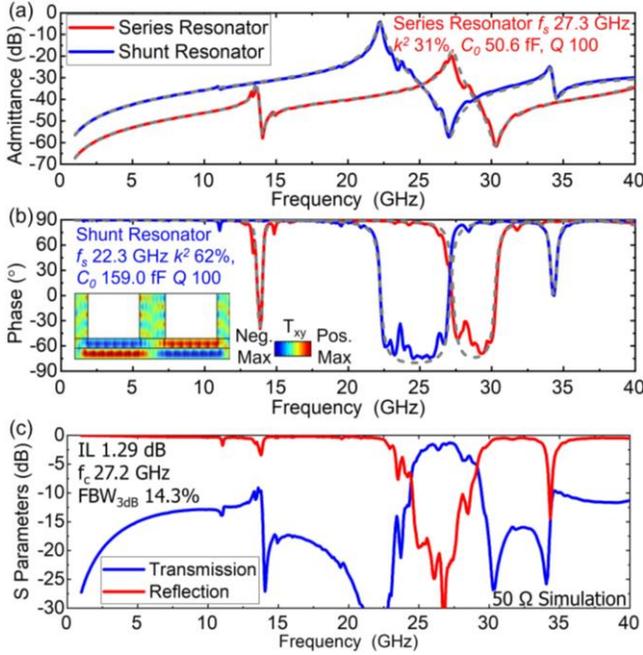

Fig. 4 Simulated admittance (a) amplitude and (b) phase, inset key specifications and vibration mode shape. (c) Simulated filter response.

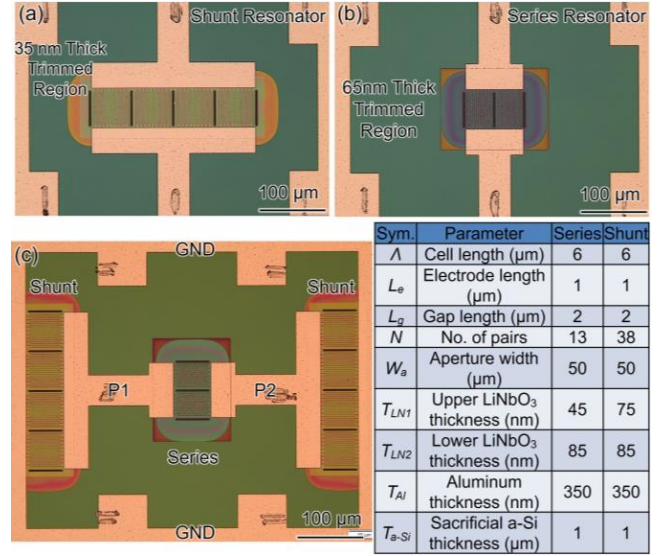

Fig. 5 Microscopic images of fabricated (a) shunt resonator, (b) series resonator, and (c) filter. Key dimensions are listed in the table.

thermal nonlinearity is severe [37], as the thin membrane has a large thermal resistance. Second, the footprints are large, as the lateral field excited structure has low capacitance density in thin films. The long traces, no longer electrically short, introduce undesired EM effects impacting filter performance [38]. Third, the loss in resonators is still high, marked by moderate quality factor ($Q$), intrinsically from the surface damping caused by the large surface-to-volume ratio [39], [40].

In this work, we demonstrate the first mmWave acoustic filters using 150 nm bi-layer periodically poled piezoelectric film (P3F) LiNbO₃ (Fig. 2), achieving low IL of 1.52 dB and 3-dB FBW of 19.4%, surpassing SoA (Fig. 1). The P3F structure promises better $Q$, smaller footprints, and better linearity. Upon further development, the reported P3F LiNbO₃ platform is promising for compact filters at mmWave.

## II. DESIGN AND SIMULATION

The third-order ladder filter consists of one series resonator and two shunt resonators [Fig. 2 (a)] [38]. The resonators comprise 350 nm thick aluminum (Al) interdigital transducers (IDT) on the top of 150 nm thick bi-layer 128° Y-cut LiNbO₃ thin-film, suspended over a sapphire substrate with 1 μm thick a-Si bonding and sacrificial layer [Fig. 2 (b)]. It is also reported that the a-Si layer helps preserve the quality of the transferred thin-film [41]. The alternating lateral electrical fields between IDTs excite the second-order symmetric (S2) mode [Fig. 3 (b)], via the piezoelectric coefficient e₁₅ in 128° Y-cut LiNbO₃. The thick electrodes on the top reduce the resistive loss and thermal resistance, while not mechanically loading the resonance as it is in the stress nodes [41]. The devices are fully anchored for higher structural strength during the release. A list of key design parameters is labeled in Fig. 2 (c) and listed in Fig. 5.

The key innovation of the work is the successful usage of the P3F LiNbO₃ stack for filters [Fig. 3 (a)], where piezoelectric thin films with opposite orientations, rotated about the axis defined by the intersection of the plane joining the two layers, are placed on top of each other, enabling higher-order mode operation in thicker films without losing electromechanical coupling ($k^2$) [11], [20], [42]–[45]. Specifically, the thickness of each layer matches the half wavelength in the higher-order thickness modes [Fig. 3 (a)]. Thus, piezoelectrically generated charges build up and can be effectively picked up by a single transducer on the top of the P3F stack. The P3F LiNbO₃ platform has shown promising experimental results for high-frequency operations up to 50 GHz [11], but so far, it has not been demonstrated as filters. This work aims to demonstrate the prototype. More specifically, we are using bi-layer P3F LiNbO₃ operating in S2 mode [Fig. 3 (b)], which can be intuitively thought of as bonding two A1 mode resonators in single-layer LiNbO₃ toward a thicker film, while achieving the advantages of higher $Q$, smaller footprint, and lower thermal resistance.

COMSOL finite element analysis (FEA) simulated S2 resonator admittance is plotted in Fig. 4 (a)-(b), showing a high $k^2$ of 62% for the shunt resonator at 22.3 GHz and $k^2$ of 31% for the series resonator at 27.3 GHz. This is obtained via a modified Butterworth-Van Dyke (MBVD) fitting with multiple motional branches [Fig. 6 (c)] [46]. Series inductor $L_s$ and resistors $R_s$ are ignored for FEA, as the EM effects are not coupled in COMSOL. The series and shunt resonators are designed with different film thicknesses (more specifically, 45 nm on top, 85 nm on bottom for series, and 75 nm on top, 85 nm on bottom for shunt) to achieve the necessary shift in resonance ($f_s$) [47]. The static capacitance $C_0$ is chosen to match the impedance of the filter to 50 Ω while providing 10 dB out-of-band (OoB) rejection. The resonator $Q$ is assumed to be 100 conservatively which is lower than the previous LiNbO$_3$ P3F [44]. The simulated filter [Fig. 4 (c)] shows an IL of 1.29 dB, a center frequency ($f_c$) of 27.2 GHz, a 3-dB FBW of 14.3%, and an OoB of 10.1 dB. The low-loss and wideband response promise mmWave compact filter platform using P3F LiNbO$_3$.

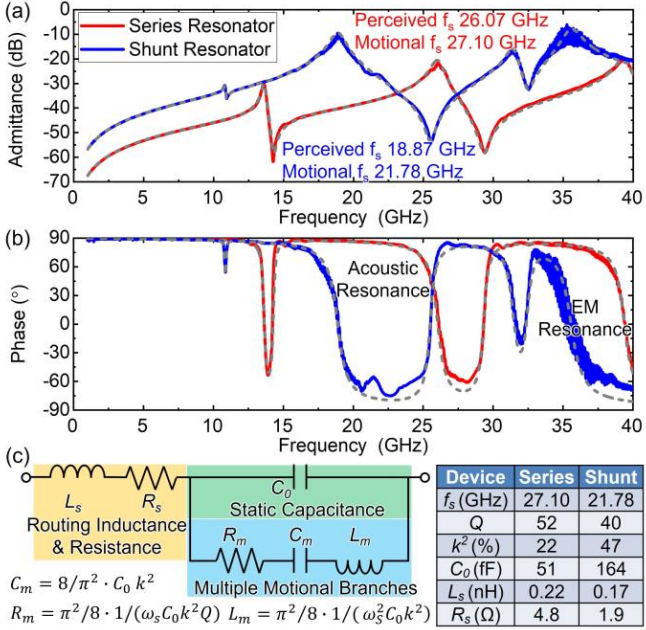

Fig. 6 Measured wideband admittance response in (a) amplitude and (b) phase. (c) Modified mmWave MBVD model and extracted key resonator specifications.

### III. FABRICATION AND MEASUREMENT

The stack for this work is provided by NGK Insulators Ltd. The process starts by patterning the resonator regions, then trimming the LiNbO$_3$ thickness of the active regions to the desired values using ion beam-assisted argon gas cluster trimming, which is reported to maintain surface roughness and high crystallinity through material analysis [48]. Next, the top electrodes are patterned with 350 nm evaporated Al IDTs. Additionally, the probing pad and routing regions are thickened up by another 350 nm evaporated Al to lower the resistive loss. Afterward, release windows are defined and etched using the ion beam. The resonator bank is divided into multiple resonators for easier release of the filters. Finally, metal electrodes are patterned, and the resonators are released through xenon difluoride (XeF$_2$) Si etch.

Optical images of the fabricated stand-alone shunt and series resonators and the filter are displayed in Fig. 5 (a), (b), and (c). The key dimensions are listed in Fig 5. The filter has a small footprint of 0.85 mm by 0.75 mm.

The resonators and the filter are first measured using a Keysight vector network analyzer (VNA) in the air at a −15 dBm power level. The admittance response in amplitude and phase of the resonators are plotted in Fig. 6 (a) and (b), fitted with the mmWave MBVD circuit model [41] in Fig. 6 (c).

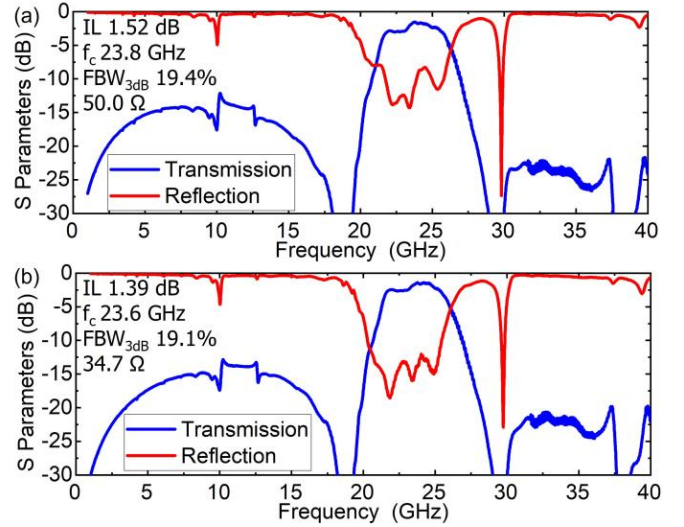

Fig. 7 Measured filter wideband transmission and reflection with (a) 50 Ω and (b) matched to 34.7 Ω port impedance.

A higher frequency resonance of electromagnetic (EM) nature occurs due to the self-resonance of the reactive parasitic embedded in the resonator routing. $L_s$ and routing resistance $R_s$ are fitted based on EM resonances. Another effect is that the perceived resonances are now at 26.07 GHz for the series and 18.87 GHz for the shunt, which deviate from the mechanical resonances represented by the motional elements $L_m$, $C_m$, and $R_m$. The measurements show $Q$ around 40 and a high $k^2$ of around 47% (Fig. 6 inset table). $k^2$ is obtained via MBVD fitting, which is smaller than the perceived $k^2$ calculated from $f_s$ and $f_p$, due to the existence of Ls [41]. $Q$ is defined as the anti-resonance due to the inclusion of $R_s$ and $L_s$.

The measured filter response under 50 Ω is shown in Fig. 7 (a). The 23.8 GHz filter exhibits a low IL of 1.52 dB IL, a wide 3-dB FBW of 19.4%, a 30 dB Shape Factor of 2.08, and an OoB rejection of 12.1 dB, matching device simulation. The measured filter response under resistance matched 34.7 Ω in Fig. 7 (b) shows a passband centered at 23.6 GHz with IL of 1.39 dB, broad 3-dB FBW of 19.1%, OoB rejection of 12.8 dB and 30 dB Shape Factor of 2.16. Compared with SoA low-loss acoustic filters (Fig. 1), this work shows significant frequency scaling and FBW enhancement. The reduced OoB performance is a drawback of using a low order filter with only three resonators and will be improved in future works.

### IV. CONCLUSION

In this work, we demonstrate the first mmWave acoustic filters using 150 nm bi-layer P3F LiNbO$_3$, achieving low IL of 1.52 dB and 3-dB FBW of 19.4%, surpassing SoA. The P3F structure promises better $Q$, smaller footprints, and better linearity. Upon further development, the reported P3F LiNbO$_3$ platform is promising for compact filters at mmWave.

ACKNOWLEDGMENT

The authors would like to thank the funding support from the DARPA COFFEE program and Dr. Ben Griffin and Dr. Todd Bauer for helpful discussions.